\documentclass[a4paper,11pt]{article}
\usepackage{graphicx}
\usepackage{caption}
\usepackage{subcaption}
\usepackage{amssymb}
\begin{document}
\title{A new expansion of the Coulomb potential and linear $r_{ij}$ terms}
\author{Richard Habrovsk\'y\\
Jaroslav Heyrovsky Institute of Physical chemistry, Czech Academy of Sciences,\\
 Dolej\v{s}kova 3, CZ-18200 Prague, Czech Republic\\
email: richard.habrovsky@jh-inst.cas.cz\\}

\maketitle

\begin{abstract}

In this work a new expansions of Coulomb potential and interparticle distance - linear term $r_{12}$
are proposed. Except the singularities, the expansions converge to the exact value in the whole
coordinate space including the vicinity of the singularities or correlation cusp points.
The disadvantage of the expansion is, that it leads to complicated goniometric functional forms. 
Anyway, for their simplification, we used the method of their expansion developed by the author in the past.
\end{abstract}

\section{Introduction}
\label{}
%\LaTeX Since the beginning of the Quantum Physics, it is well known \cite{RefHyll1,RefHyll2}, that one of the necessary conditions
% ZLEPSI AJ!!!
Since the beginning of the quantum physics, it is well known \cite{RefHyll1,RefHyll2} that one 
of the necessary conditions of a rapid convergence towards the exact values of energy and
other properties of systems consisting of few to many
 particles is inclusion of the interparticle coordinates $r_{ij}$ into the wave function. In previous work \cite{RefHabro} the author has shown,
 that choosing special  curvilinear coordinates together with inclusion of $r_{12}$ powers, leads to results for 3-body problem with very high
 precision. This approach was tested on helium and hydrogen anion system. There were also proposed exact (in the sense of classical 
 quantum mechanics) integro-differential equations for determination of basis set of Helium like ions.
 Anyway, extension of different variational approaches with $r_{ij}$ coordinates leads to serious problems
 with the integration of overlap or Hamiltonian matrices,that in many cases were overcome by different expansions 
 of linear $r_{ij}$ and Coulomb $1/r_{ij}$ terms. These approximations caused nonnegligible errors in energy and other atomic/molecular properties.
 A simple example of general function was proposed by Frost \cite{RefFrost} \\
% ZLEPSI AJ!!!
\begin{eqnarray}
% {\displaystyle \Phi_{i,j,k,l}=\sum_{m=0}^{\infty}\sum_{o=0}^{\infty}\prod_{i}^{n}e^{-\alpha_{i} r_{i}} \prod_{j}^{n} r_{j}^{m} \prod_{k<l} r_{kl}^{o}}, 
% {\displaystyle \Phi=\sum_{s}C_{s}\phi_{s}}, 
	\Phi=\sum_{s}C_{s}\phi_{s}, 
\end{eqnarray}
where
\begin{eqnarray}
	\phi_{s}=\prod_{i<j}\phi_{ij}^{(s)},\\
	\phi_{ij}^{(s)}=r_{ij}^{n_{ij}}\exp{(-\zeta_{ij}r_{ij})}.
	%ZLE : \phi_{ij}^{s}=r_{ij}^{n_{ij}}\exp{(-\zeta_{ij}r_{ij})}.
\end{eqnarray}
 % and last product contains all combinations of the electron pairs $r_{kl}$,which are equal to  ${n \choose 2}$ .   
The complexity of approaches using $r_{ij}$ terms increases exponentially with the number of particles in the studied system.
Different methods were proposed, that  avoid this problem. As an examples of such approaches we can mention CCR12/CCR12F12 methods(\cite{RefCCR12},
\cite{RefCCR12-F12-1},\cite{RefCCR12-F12-2}), or the work \cite{RefCoulPT2}. Anyway, we think, that the principal way how to solve this problem 
is expansion of $r_{ij}$ powers (positive or negative) into the power series (\cite{RefGill},\cite{RefCohl},\cite{RefAng}).
Remarkable work was done by Gill \cite{RefGill}. He proposed a method which correctly describes the tail of the Coulomb potential, but the solution for 
the singular part of the potential was still open. 
% SPOMENUT Nas clanok v JCP)
% SPOMENUT R12-F12 metody v PT a CC teoriach ( Kuttzelnig, Noga pripadne Wim Klopper)
In this work we propose an expansion of the Coulomb potential and the linear $r_{ij}$ term, which converges in the whole space except
at the singularity itself. The price to pay for this is the complexity of the function of individual terms. Here we must realize, that the output 
of this method are three dimensional complicated goniometric functions. To overcome the problems with the final integrals, these functions must be expanded 
by another expansions into (integreable) goniometric functions.

\section{Derivation of the new expansions of the linear $r_{12}$ term and Coulomb potential}

Our aime is to express $r_{12}$ and $1/r_{12}$ in a form which would be suitable for evaluation 
of integrals needed in many body problems. Using elementary trigonometry we have
%We start with the definition of the linear $r_{12}$ term in spherical coordinates (cosine Lemma) 
%\newpage
%
\begin{eqnarray}
&r_{12}=&
\sqrt{r_{1}^{2}+r_{2}^{2}-2r_{1}r_{2}\cos(\theta_{12})}
\label{Def_r12}
\end{eqnarray}
where $\theta_{12}$ is the angle between vectors $\bf{r_{1}}$ and $\bf{r_{2}}$, $\cos(\theta_{12})=\bf{r_{1}}.\bf{r_{2}}/(|r_{1}||r_{2}|)$. In spherical 
coordinates (see. Fig1) we obtain for $\cos(\theta_{12})$

\begin{eqnarray}
\cos(\theta_{12})=\cos(\theta_{1})\cos(\theta_{2})+\sin(\theta_{1})\sin(\theta_{2})\cos(\phi_{1}-\phi_{2})
\label{Def_Cos}
\end{eqnarray}

% TO DOOOOOOO!
\begin{figure}
%Fig1
\centering
	\caption{The vectors $\vec{r_{1}}$ and $\vec{r_{2}}$ in the spherical coordinates} 
	\includegraphics[width=0.7\linewidth]{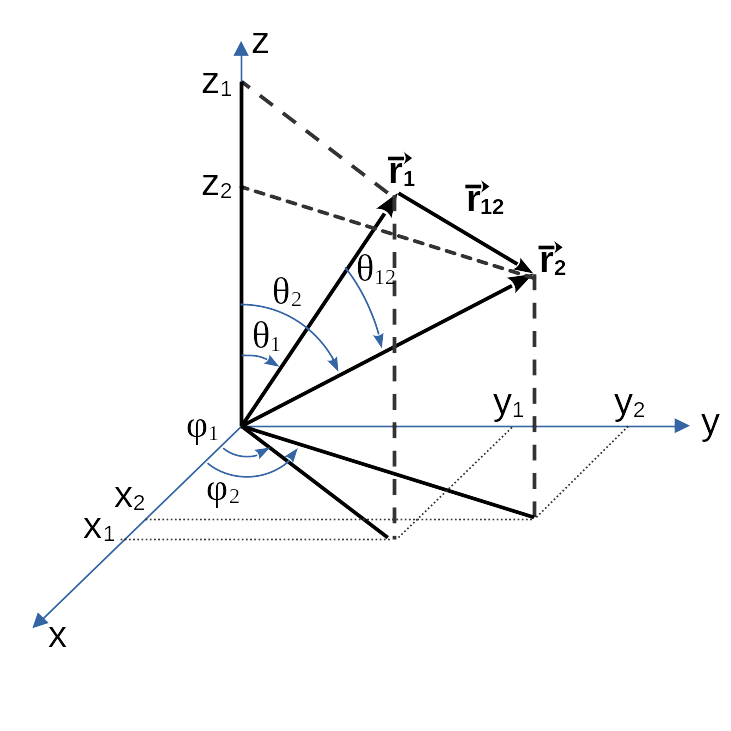}
\end{figure}

% TO DOOOOOOO!
%
It is clear that a direct Taylor expansion of $r_{12}$ or $1/r_{12}$ term, does not converge in the whole space,see Gill \cite{RefGill}.
An alternative is an expansion employing the terms $r_{<}^{l}/r_{>}^{(l+1)}\mathit{P_{l}}(\cos{(\theta_{12})})$, 
where $\mathit{P_{l}}(\cos{(\theta_{12})})$ represents Legendre polynomials \cite{RefBethe}. Unfortunatelly 
this expansion does not converge everywhere, especially  at the larger vicinity of singular points. \nonumber\\
Now we show an alternative method of expansion of the linear $r_{12}$ function  and its Coulombic counterpart $1/r_{12}$ . Let us assign 
$r_{>}$ and $r_{<}$ to $r_{1}$ and $r_{2}$ in such a way that $r_{>} \geq r_{<}$. We employ the Goldman coordinates \cite{RefGoldm}

\begin{eqnarray}
u=r_{>},  v=\frac{r_{<}}{r_{>}}
\label{Goldcoord}
\end{eqnarray}
Using (\ref{Goldcoord}) we can now write

\begin{eqnarray}
r_{12}=u\sqrt{1-2.v.\cos{\theta_{12}}+v^{2}}=u\sqrt{1-2.v.\Big(\cos^{2}(\frac{\theta_{12}}{2})-\sin^{2}(\frac{\theta_{12}}{2})\Big)+v^{2}}\
\end{eqnarray}

Regrouping the terms we obtain the following form 
\begin{eqnarray}
r_{12}=u\sqrt{\sin^{2}(\frac{\theta_{12}}{2})(1+v)^{2}+\cos^{2}(\frac{\theta_{12}}{2})(1-v)^{2}}\
\label{r12SinpolCospol}
\end{eqnarray}

Notice that $\sin(\theta_{12}/2)$ and $\cos(\theta_{12}/2)$ are always non-negative in our case since $0\le\theta_{12}\le \pi$.
Moreover, from (\ref{Goldcoord}) immediately follows $0\le v\le 1$, which guarantees

\begin{eqnarray}
(1+v)\sin(\frac{\theta_{12}}{2})\ge0
\end{eqnarray}

\begin{eqnarray}
(1-v)\cos(\frac{\theta_{12}}{2})\ge0. 
\end{eqnarray}

%Next we propose the following approach to tackle the singularities in the Coulomb potential:
%we can compare the terms $(1+v)\sin(\frac{\theta_{12}}{2})$ and $(1-v)\cos(\frac{\theta_{12}}{2})$, 
%and define two subspaces of 2-dimensions $(v,\theta_{12})\in \langle 0,1 \rangle \times \langle 0,\pi \rangle$  \nonumber\\ 

Now we can define two regions in $(v,\theta_{12})$-coordinate system (see Fig.2)

subspace A\\
\begin{eqnarray}
{(1+v)\sin(\frac{\theta_{12}}{2})}\le{(1-v)\cos(\frac{\theta_{12}}{2}))}
\label{Ineq1}
\end{eqnarray}
subspace B elsewhere.

%We also introduce functions $X$ and $S$:
In $(v,\theta_{12})$-coordinates we introduce the functions $X$ and $S$ in the following way:\\
 for the A subspace\\ 
\begin{eqnarray}
X=\tan(\frac{\theta_{12}}{2})\Big(\frac{1+v}{1-v}\Big),
\end{eqnarray}

while for the B subspace\\ 
\begin{eqnarray}
X=\cot(\frac{\theta_{12}}{2})\Big(\frac{1-v}{1+v}\Big),
\end{eqnarray}

the function $S$ in the A subspace we define as

\begin{eqnarray}
S=u.\cos(\frac{\theta_{12}}{2})(1-v)
\end{eqnarray}
while in the B subspace 

\begin{eqnarray}
S=u.\sin(\frac{\theta_{12}}{2})(1+v)
\end{eqnarray}

%It is easy to see that the Coulomb potential can be expressed in terms of the functions S and X in both 
It is easy to see that the Coulomb potential can be expressed in both regions A,B in the same way, 

\begin{eqnarray}
\frac{1}{r_{12}}=\frac{1}{S\sqrt{1+X^{2}}}
\end{eqnarray}

\begin{figure}
\centering
%\caption{Definition of the subspaces A,B and C} !!!!! TO DO REMOVE C!!!
\caption{Definition of the subspaces A,B}
\includegraphics[width=0.7\linewidth]{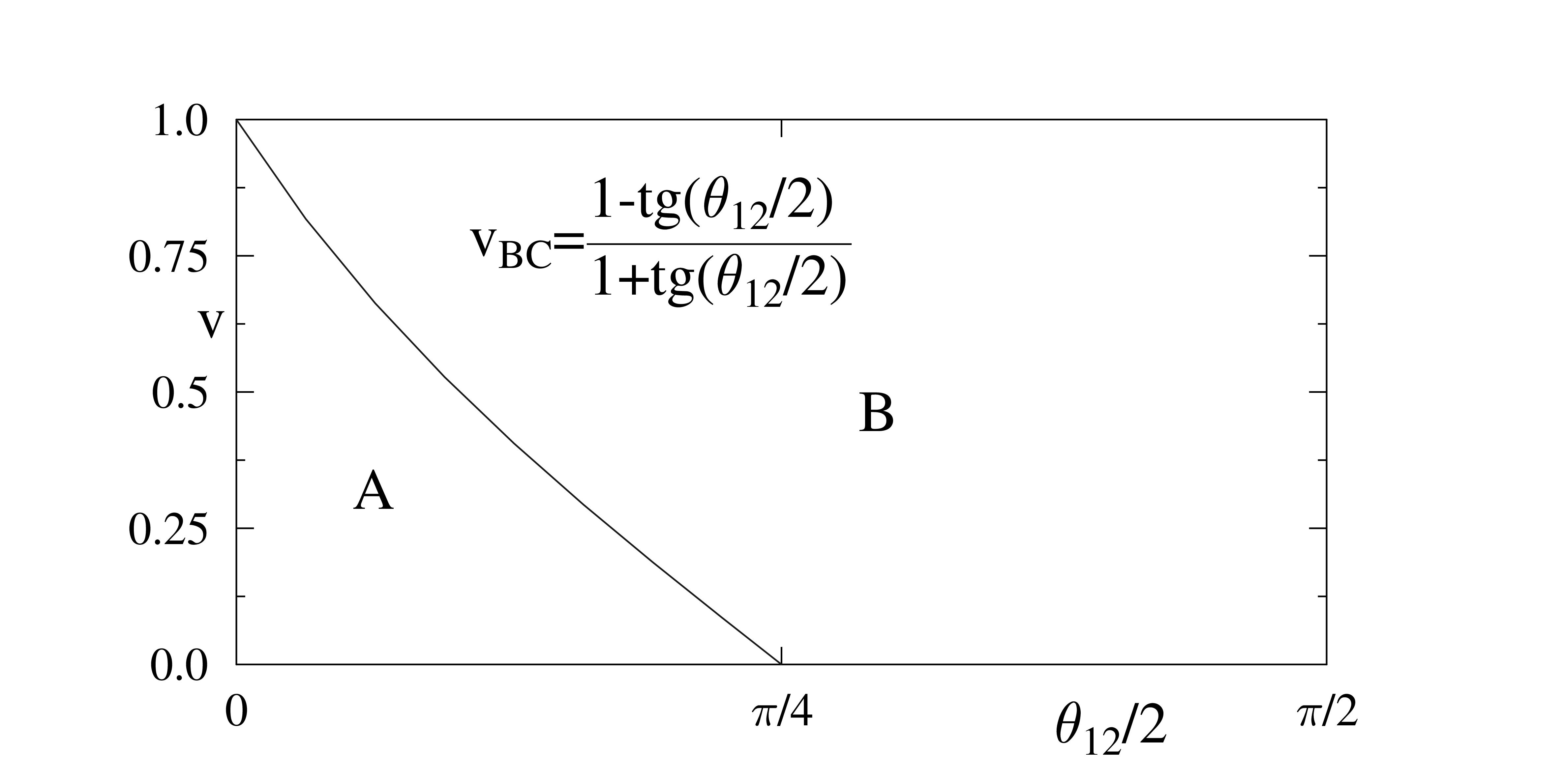}
\end{figure}

For the boundary between A and B  (Fig.2) follows,

\begin{eqnarray}
v_{BC}=\frac{1-\tan(\frac{\theta_{12}}{2})}{1+\tan(\frac{\theta_{12}}{2})}
\end{eqnarray}

%\begin{eqnarray}
%TEST: \sin^{2}(\theta_{-})
%\end{eqnarray}

%Here we must take into account that $\sin(\theta_{12}/2)$ and $\cos(\theta_{12}/2)$ are always
%positive because $\frac{\theta_{12}}{2} \in \langle 0,\frac{\pi}{2} \rangle$. 
%From $\sin(\frac{\theta_{12}}{2})\ge\cos(\frac{\theta_{12}}{2})$ for the interval $\frac{\theta_{12}}{2} \in \langle \frac{\pi}{4},\frac{\pi}{2} \rangle$
%and $(1+v)\ge(1-v)$ is clear, that the functions $S$ and $X$  for the subspace-B see (Fig.2).

%Hence the integration through $v$ coordinate splits in these parts:
%For A subspace: $\int\limits_{v_{BC}}^{1}Function_{1}dv$:q!
%For B subspace: $\int\limits_{0}^{v_{BC}}Function_{2}dv$

%We must underline that the function X was defined on the subspaces A and B in such a way to satisfy the condition $X\le1$. This guarantees 
On both regions we now have $X\le1$. This guarantees the convergence  of the Taylor expansion of the functions $\sqrt{1+X^2}$ (in the linear term)
and $\frac{1}{\sqrt{1+X^2}}$ (in the Coulomb potential).
Provided $X\le1$ these functions can be expanded at the point $X^{2}=X_{c}$:

\begin{eqnarray}
r_{12}=S.\sqrt{1+X^{2}}=S.\Big(\sqrt{1+X_{c}}+\frac{1}{2}\frac{(X^{2}-X_{c})}{(1+X_{c})^\frac{1}{2}}-\frac{1.1}{2.4}\frac{(X^{2}-X_{c})^{2}}{(1+X_{c})^\frac{3}{2}}+\nonumber\\ 
+\frac{1.1.3}{2.4.6}\frac{(X^{2}-X_{c})^{3}}{(1+X_{c})^\frac{5}{2}}-\frac{1.1.3.5}{2.4.6.8}\frac{(X^{2}-X_{c})^{4}}{(1+X_{c})^\frac{7}{2}}+\dots\Big)
\label{Expansion1}
\end{eqnarray}

the square root in the denominator can be expanded similarly 
\begin{eqnarray}
\frac{1}{r_{12}}=\frac{1}{S.\sqrt{1+X^{2}}}=\frac{1}{S}\Big(\frac{1}{\sqrt{1+X_{c}}}-\frac{1}{2}\frac{(X^{2}-X_{c})}{(1+X_{c})^\frac{3}{2}}+\frac{1.3}{2.4}\frac{(X^{2}-X_{c})^{2}}{(1+X_{c})^\frac{5}{2}}-\nonumber\\
\frac{1.3.5}{2.4.6}\frac{(X^{2}-X_{c})^{3}}{(1+X_{c})^\frac{7}{2}}+\frac{1.3.5.7}{2.4.6.8}\frac{(X^{2}-X_{c})^{4}}{(1+X_{c})^\frac{9}{2}}-\dots\Big)=\nonumber\\
\frac{1}{\sqrt{1+X_{c}}}+\sum_{k=1}^{\infty}(-1)^{k}\frac{(2k-1)!!}{2k!!}\frac{(X^{2}-X_{c})^{k}}{(1+X_{c})^\frac{2k+1}{2}}
\label{Coul_Expan1} 
\end{eqnarray}

It is now usefull to define for (\ref{Expansion1}) the coefficients 
\begin{eqnarray}
	C_{1}=\frac{1}{2},C_{2}=\frac{1.1}{2.4},C_{3}=\frac{1.1.3}{2.4.6},C_{4}=\frac{1.1.3.5}{2.4.6.8},...\\
\label{Coeff1} 
\end{eqnarray}
and for (\ref{Coul_Expan1}) the coefficients 
\begin{eqnarray}
	D_{k}=\frac{(2k-1)!!}{(2k!!)}.
\label{Coeff2} 
\end{eqnarray}
%After expansion of denumerator of $X$ and $S$  we obtained an expansion of the linear function $r_{12}$ 
%and the Coulomb potential with separate functions of $u$, $v$, and $\frac{theta_{12}}{2}$. This is the first step of the method.
%Now we must realize that direct usage of this method leads for many particles integrals with goniometric functions 
%$\sin(\frac{\theta_{ij}}{2})=\sqrt{\frac{1-\cos(\theta_{ij})}{2}}$ and $\cos(\frac{\theta_{ij}}{2})=\sqrt{\frac{1+\cos(\theta_{ij})}{2}}$ to large problems.

%Second, we expanded the goniometric functions with the half-angle argument with the method utilized by author for a long time. 
%Due to extensity of the work, the method will be published in the near future.

After expansion of the denumerator of $X$ and $S$  we obtaine an expansion of the linear function $r_{12}$ 
and the Coulomb potential with separate functions of $u$, $v$, and $\theta_{12}/2$. This is the first step of the method.
Now we must realize that direct usage of this expansion leads for many particle integrals to almost intractable problems.
It is easy to see it from the type of the integrals with our functions $\sin{\frac{\theta_{ij}}{2}}=\sqrt{\frac{1-\cos{\theta_{ij}}}{2}}$ and 
$\cos{\frac{\theta_{ij}}{2}}=\sqrt{\frac{1+\cos{\theta_{ij}}}{2}}$. The integrations with inclusion of such functions with different combination of angles
$\theta_{ij}/2$ lead to large problems, mainly to integrations with combinations of odd powers of our basic goniometric functions.
Second, we expanded the goniometric functions with the half-angle argument with the method utilized by author for a long time. 
In the past author developed a package for expansion of goniometric functions occured in the expansions of the functions
 $S\sqrt{1+X^{2}}$ and $\frac{1}{S\sqrt{1+X^{2}}}$.
It is clear that direct expansion of positive and negative powers of functions $\sin{\frac{\theta_{12}}{2}}=\sqrt{\frac{1-\cos{\theta_{12}}}{2}}$
 at the point $\theta_{12}=0$ and in its very near surrounding and simiraly for $\cos{\frac{\theta_{12}}{2}}=\sqrt{\frac{1+\cos{\theta_{12}}}{2}}$
at the point $\theta_{12}=\pi$ leads to large errors in expansion series.\\
To overcome these problems we propose the following method for expansion of goniometric parts of our expansion of the Coulomb 
potential and the linear term. First we define new angular coordinates by these two regular transformations (Jacobian of the transformation 
is non-zero at the whole domain)
\begin{eqnarray}
\theta_{+}=\frac{\theta_{1}+\theta_{2}}{2}, \theta_{-}=\frac{\theta_{1}-\theta_{2}}{2}
\label{Regulth1}
\end{eqnarray}
\begin{eqnarray}
\phi_{+}=\frac{\phi_{1}+\phi_{2}}{2},\phi_{-}=\frac{\phi_{1}-\phi_{2}}{2}.
\label{Regulphi2}
\end{eqnarray}
 Fig.3  shows the Domain1 of the regular transformation (\ref{Regulth1}) and the Domain2 of (\ref{Regulphi2}), that could be used as subintegration domains.
In the Domain1 the y axes belongs to $\theta_{+}/2$ and x axes belongs to $\theta_{-}/2$. Here the border is defined by $\theta_{+}/2$
which is linear function of  $\theta_{-}/2$. Similar statements are valid for Domain2 with exchange $\theta \leftrightarrow \phi$.

% Fig.1 Fig.2%
%\begin{table*} 
\begin{figure}%
    \centering
    \begin{subfigure}[b]{0.4\textwidth}
%   \begin{subfigure}[b]{0.3\textwidth}
        \centering
        \includegraphics[width=\textwidth]{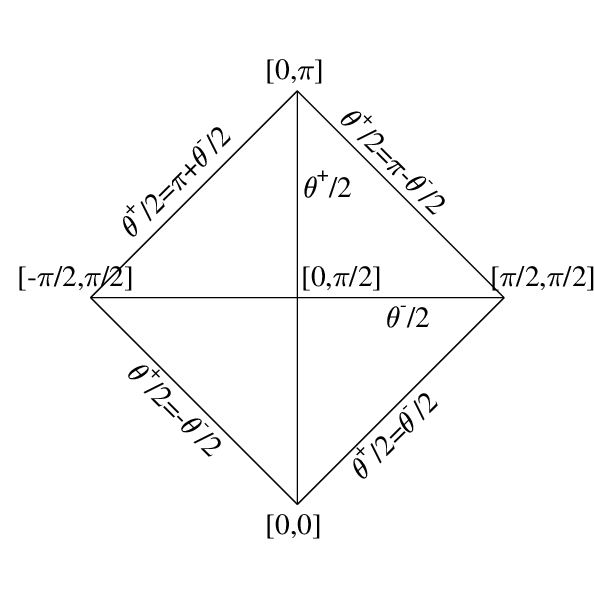}
	\caption{Domain 1}
%       \label{Fig.2}
    \end{subfigure}
    \hfill   
    \begin{subfigure}[b]{0.4\textwidth}
%   \begin{subfigure}[b]{0.7\textwidth}
        \centering
        \includegraphics[width=\textwidth]{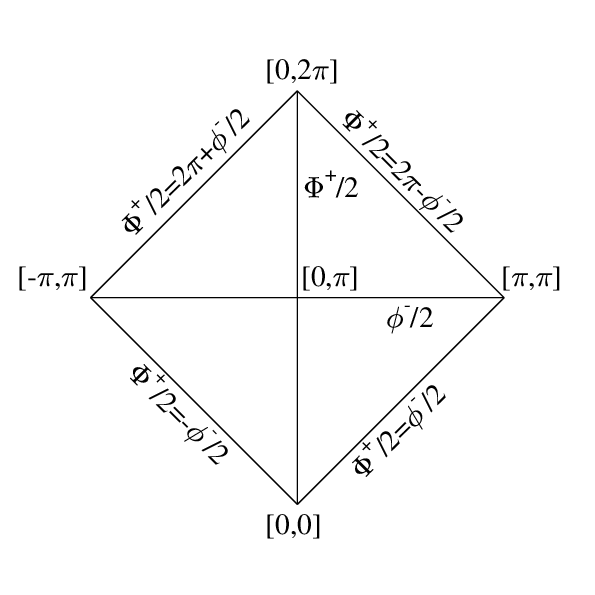}
	\caption{Domain 2}
%       \label{Fig.3}
    \end{subfigure}
    \caption{Domains after regular transformations}
%    \label{Domains after regular transformations}
\end{figure}
%    \subfloat[\centering label 2]{\includegraphics[width=5cm]{./SubSpace.eps} }%
%    \caption{ Domains after regular transformations}%
%    \label{Fig: Domain}%
%\includegraphics[scale=0.75,natwidth=640,natheight=480]{./SubSpace2.eps}
% &\begin{figure}
%\includegraphics[scale=0.3,natwidth=640,natheight=480]{./SubSpace.eps} 
%\end{figure}
%\hline
%\end{tabular}
%\end{table*}

%%\begin{figure}
%%\centering
%%\caption{ Domain1}
%%\includegraphics[width=0.7\linewidth]{./SubSpace2.eps}

%\includegraphics[width=0.7\linewidth]{./DivSpace_white_bkGrnd.png}
%\includegraphics[width=\textwidth]{/home/richheard1112/CLANOK3/NEW_AR12T/2dFigures/DivSpace_white_bkGrnd.png}
%%\end{figure}
%%%\begin{figure}
%%\centering
%%\caption{ Domain2}
%\includegraphics[width=0.7\linewidth]{./SubSpace.eps}
%\includegraphics[width=0.7\linewidth]{./DivSpace_white_bkGrnd.png}
%\includegraphics[width=\textwidth]{/home/richheard1112/CLANOK3/NEW_AR12T/2dFigures/DivSpace_white_bkGrnd.png}
%%\end{figure}
%\begin{figure}[ht]
%\caption{\tiny Fig.1.: Definition of subspaces A,B and C}
%\includegraphics{/home/richheard1112/CLANOK3/NEW_AR12T/2dFigures/DivSpace_white_bkGrnd.png}
%\includegraphics[width=\textwidth]{/home/richheard1112/CLANOK3/NEW_AR12T/2dFigures/SubSpace_white_bkGrnd.png}
%\end{figure}

It is easy to prove the following identities

\begin{eqnarray}
\sin{\frac{\theta_{12}}{2}}=\sqrt{\sin^{2}(\theta_{-})\cos^{2}(\phi_{-})+\sin^{2}(\theta_{+})\sin^{2}(\phi_{-})}
\label{DefSinpol}
\end{eqnarray}

\begin{eqnarray}
\cos{\frac{\theta_{12}}{2}}=\sqrt{\cos^{2}(\theta_{-})\cos^{2}(\phi_{-})+\cos^{2}(\theta_{+})\sin^{2}(\phi_{-})}.
\label{DefCospol}
\end{eqnarray}
%\cos{\frac{\theta_{12}}{2}}=\sqrt{\cos{\theta_{-}}^{2}\cos{\phi_{-}}^{2}+\cos{\theta_{+}}^{2}\sin{\phi_{-}}^{2}}

We can treat equations (\ref{DefSinpol}) and (\ref{DefCospol}) in a similar way as we did for \ref{Ineq1} for the subspace A and its complement B\\

\begin{eqnarray}
	\rm{Subspace} & \rm{C}: |\sin{\theta_{-}}||\cos{\phi_{-}}|\ge|\sin{\theta_{+}}||\sin{\phi_{-}}| \\
	\rm{Subspace} & \rm{D}: |\sin{\theta_{-}}||\cos{\phi_{-}}|\le|\sin{\theta_{+}}||\sin{\phi_{-}}|. 
\end{eqnarray}

It is necessary to underline independence of functions $\cos(\theta_{12})$, $\sin(\theta_{12}/2)$ and $\cos(\theta_{12}/2)$ on the angle $\phi_{+}$,
so definition of subspaces does not depend on $\phi_{+}$ either. Upper inequalities define $\phi_{-}$ as two dimensional function, that determines subspaces, 
where the expansions of the goniometric functions converge

\begin{eqnarray}
|\tan{\phi_{-}}|=\frac{|\sin{\theta_{-}}|}{|\sin{\theta_{+}}|} \\
\end{eqnarray}

Defining $W_{1}$ and $W_{2}$ functions
\begin{eqnarray}
W_{1}=\frac{|\sin{\theta_{+}}|}{|\sin{\theta_{-}}|}|\tan{\phi_{-}}|\\
W_{2}=\frac{|\sin{\theta_{-}}|}{|\sin{\theta_{+}}|}|\cot{\phi_{-}}|
\end{eqnarray}
we can rewrite (\ref{DefSinpol}) separetely on the subspaces as 

\begin{eqnarray}
	\rm{Subspace}&\rm{C}: \sin{\frac{\theta_{12}}{2}}=|\sin{\theta_{-}}||\cos{\phi_{-}}|\sqrt{1+W_{1}^{2}}\\
	\rm{Subspace}&\rm{D}: \sin{\frac{\theta_{12}}{2}}=|\sin{\theta_{+}}||\sin{\phi_{-}}|\sqrt{1+W_{2}^{2}}
\end{eqnarray}

Analogous manipulations can be done for \ref{DefCospol}, by changing $\sin{\theta_{-}}$ to 
$\cos{\theta_{-}}$ and $\sin{\theta_{+}}$ to $\cos{\theta_{+}}$.  

\begin{eqnarray}
W_{3}=\frac{|\cos{\theta_{+}}|}{|\cos{\theta_{-}}|}|\tan{\phi_{-}}|\\
W_{4}=\frac{|\cos{\theta_{-}}|}{|\cos{\theta_{+}}|}|\cot{\phi_{-}}|
\end{eqnarray}
we can rewrite (\ref{DefSinpol}) separetely on the subspaces as 

\begin{eqnarray}
	\rm{Subspace}&\rm{E}: \cos{\frac{\theta_{12}}{2}}=|\cos{\theta_{-}}||\cos{\phi_{-}}|\sqrt{1+W_{3}^{2}}\\
	\rm{Subspace}&\rm{F}: \cos{\frac{\theta_{12}}{2}}=|\cos{\theta_{+}}||\sin{\phi_{-}}|\sqrt{1+W_{4}^{2}}.
\end{eqnarray}
It is necessary to introduce expansions of $X(Space)$ and $1/S(Space)$ (the X and S functions depend on Subspaces A or B) 

\begin{eqnarray}
	\tilde{X}(A)=\frac{(1+t)}{(1-t)}.R_{sin}.R_{cos^{-1}}\\
	\tilde{X}(B)=\frac{(1-t)}{(1+t)}.R_{cos}.R_{sin^{-1}},
\end{eqnarray}
\begin{eqnarray}
	\frac{1}{\tilde{S}(A)}=\frac{1}{(1-t)}.R_{cos^{-1}}\\
	\frac{1}{\tilde{S}(B)}=\frac{1}{(1+t)}.R_{sin^{-1}},
\end{eqnarray}

where $R_{sin/cos/sin^{-1}/cos^{-1}}$  are purely goniometric functions, that depend on subspaces C,D,E,F

\begin{eqnarray}
	R_{sin}(C)=|\sin{\theta_{-}}||\cos{\phi_{-}}|.(\sqrt{1+W_{1c}}+\nonumber\\ 
	\sum_{l=1}^{N}(-1)^{l+1}C_{l}\frac{(W_{1}^{2}-W_{1c})^{l}}{(1+W_{1c})^{\frac{2l-1}{2}}}),\\
	R_{sin}(D)=|\sin{\theta_{+}}||\sin{\phi_{-}}|.(\sqrt{1+W_{2c}}+ \nonumber\\
	\sum_{l=1}^{N}(-1)^{l+1}C_{l}\frac{(W_{2}^{2}-W_{2c})^{l}}{(1+W_{2c})^{\frac{2l-1}{2}}}),
\end{eqnarray}
\begin{eqnarray}
	R_{cos}(E)=|\cos{\theta_{-}}||\cos{\phi_{-}}|.(\sqrt{1+W_{3c}}+\nonumber\\ 
	\sum_{l=1}^{N}(-1)^{l+1}C_{l}\frac{(W_{3}^{2}-W_{3c})^{l}}{(1+W_{3c})^{\frac{2l-1}{2}}}),\\
	R_{cos}(F)=|\cos{\theta_{+}}||\sin{\phi_{-}}|.(\sqrt{1+W_{4c}}+ \nonumber\\
	\sum_{l=1}^{N}(-1)^{l+1}C_{l}\frac{(W_{4}^{2}-W_{4c})^{l}}{(1+W_{4c})^{\frac{2l-1}{2}}}),
\end{eqnarray}

and for negative powers of $\sin(\theta_{12}/2)$ and $\cos(\theta_{12}/2)$ we have

\begin{eqnarray}
	R_{sin^{-1}}(C)=\frac{1}{|\sin{\theta_{-}}||\cos{\phi_{-}}|}.(\frac{1}{\sqrt{1+W_{1c}}}+\nonumber\\ 
	\sum_{l=1}^{N}(-1)^{l}D_{l}\frac{(W_{1}^{2}-W_{1c})^{l}}{(1+W_{1c})^{\frac{2l+1}{2}}}),\\
	R_{sin^{-1}}(D)=\frac{1}{|\sin{\theta_{+}}||\sin{\phi_{-}}|}.(\frac{1}{\sqrt{1+W_{2c}}}+\nonumber\\
	\sum_{l=1}^{N}(-1)^{l}D_{l}\frac{(W_{2}^{2}-W_{2c})^{l}}{(1+W_{2c})^{\frac{2l+1}{2}}}),
\end{eqnarray}
\begin{eqnarray}
	R_{cos^{-1}}(E)=\frac{1}{|\cos{\theta_{-}}||\cos{\phi_{-}}|}.(\frac{1}{\sqrt{1+W_{3c}}}+\nonumber\\ 
	\sum_{l=1}^{N}(-1)^{l}D_{l}\frac{(W_{3}^{2}-W_{3c})^{l}}{(1+W_{3c})^{\frac{2l+1}{2}}}),\\
	R_{cos^{-1}}(F)=\frac{1}{|\cos{\theta_{+}}||\sin{\phi_{-}}|}.(\frac{1}{\sqrt{1+W_{4c}}}+\nonumber\\
	\sum_{l=1}^{N}(-1)^{l}D_{l}\frac{(W_{4}^{2}-W_{4c})^{l}}{(1+W_{4c})^{\frac{2l+1}{2}}}),
\end{eqnarray}

A package was written by the author that allows us enumarate Taylor series for both spaces A and B and for positive and negative
powers of $\sin(\theta_{12}/2)$ and $\cos(\theta_{12}/2$. Now we can use  (\ref{Expansion1}) and (\ref{Coul_Expan1})
with exchange $X \leftrightarrow \tilde{X}$ and $S \leftrightarrow \tilde{S}$ to form final expansions
\begin{eqnarray}
	r_{12}\doteq\tilde{S}(\sqrt{1+X_{c}}+\sum_{l=1}^{N}(-1)^{l+1}C_{l}\frac{(\tilde{X}^{2}-X_{c})^{l}}{(1+X_{c})^{\frac{2l-1}{2}}}),\\
	\frac{1}{r_{12}}\doteq\frac{1}{\tilde{S}}(\frac{1}{\sqrt{1+X_{c}}}+\sum_{l=1}^{N}(-1)^{l}D_{l}\frac{(\tilde{X}^{2}-X_{c})^{l}}{(1+X_{c})^{\frac{2l+1}{2}}}).\\
\end{eqnarray}

%ZLEPSIT AJ!!!
%Now we are able to expand the goniometric part of all $X$ and $S$ functions. A package was written by the author that allows us enumarate 
%Taylor series for both spaces A and B and for positive and negative powers of $\sin{\frac{\theta_{12}}{2}}$ and $\cos{\frac{\theta_{12}}{2}}$
%(defined by (\ref{DefSinpol}) and (\ref{DefCospol})). These functions can be expressed in a power series of the functions $W_{1}^{2}$ and 
%$W_{2}^{2}$ in the same manner, as it was done in (\ref{Expansion1}) and (\ref{Coul_Expan1}).\\ %ZLEPSIT AJ

%%Taylor series for both (\ref{Ineq1}) (\ref{Ineq2}) and for positive and negative powers of $\sin{\frac{\theta_{12}}{2}}$ and $\cos{\frac{\theta_{12}}{2}}$ (define by (\ref{DefSinpol}) and (\ref{DefCospol})). These functions can be expressed in a power series of the functions $W_{1}^{2}$ and 

\section{Numerical results}
To simplify the evaluation of a complete Taylor expansions we define the next function
\begin{eqnarray}
	Y^{\pm{1}}\equiv(r_{12}/u)^{\pm{1}}=(\sqrt{1-2v\cos{\theta_{12}}+v^{2}})^{\pm{1}} 
\end{eqnarray}
 We used the averaged absolute differences (AAD) and the normalized averaged absolute differences (NAAD) to statistically assess total
 Taylor expansion of Coulomb potential and linear term

\begin{eqnarray}
	AAD=\frac{1}{N_{\rm{total}}}\sum_{i=1}^{N_{\rm{total}}}|Y^{\pm{1}}_{i}-\tilde{Y}^{\pm{1}}_{i}| 
\label{DefAAD}
\end{eqnarray}
\begin{eqnarray}
	NAAD=\frac{1}{N_{\rm{total}}}\sum_{i=1}^{N_{\rm{total}}}\frac{|Y^{-1}_{i}-\tilde{Y}^{-1}_{i}|}{Y^{-1}_{i}}
\label{DefNAAD}
\end{eqnarray}
where $\tilde{Y}^{\pm{1}}$ are expanded $Y^{\pm{1}}$.
The sum over $i$-indices means the sum over the individual points of the $v$,$\theta_{1}$,$\theta_{2}$,$\phi_{-}$ space.
Step for the $v$ coordinate was $0.01$, for  $\theta_{1}$ and $\theta_{2}$ was equal to $\pi/60$ and  
finally for $\phi_{-}$ was $2\pi/60$. All singular points were excluded from the statistics and 
the remaining points with finite numerical values were statistically tested. Results are very promising.
%All data for the Coulomb term (negative power $-1$ ) were selected,
 %whether were equal to infinity or in the case of $\tilde{Y}^{-1}_{i}$ if it contained NotaNumber-NaN values. 
%In the cases mentioned above-the singularities were excluded from our statistics.The remaining points with the numerical values 
%were statistically tested. 
In the TABLE1 we calculated  the AAD and the NAAD parameters for the Coulomb potential and the AAD for $r_{12}$ term. 
The normalized NAAD parameter could not be calculated for $r_{12}$ due to fact, that this function can reach also a zero values.
In the TABLE2 we focus on the region near the singularities. The $v$ coordinate ranges from $0.9$ to $0.99999999999$
with increment $9\times10^{-k}$, where $k=2,..,10$ and angles $\theta_{-}$($\theta_{-}=\theta_{1}-\theta_{2}$) and $\phi_{-}$ from $10^{-1}$,$10^{-2}$, etc. to $10^{-15}$. 
The AAD and NAAD were calculated by averaging over $\theta_1$ angles chosen from the interval $\theta_1 \in \langle 0,\pi \rangle$  with the step 
$\delta=\pi/30$ , see TABLE2. Results are remarkably precise.%IMPROVE ENGLISH

% Tab.1 %
\begin{table*}
\centering
\caption
	{ Statistical parameters AAD (\ref{DefAAD}) and NAAD (\ref{DefNAAD}) for $1/r_{12}$ and $r_{12}$ }
\begin{tabular}{|l|c|c|c|c|}
\hline
 TAB1  &  \multicolumn{2}{|c|}{ AAD } &  \multicolumn{2}{|c|}{  NAAD } \\
\hline
Series count   & N=7 & N=10 & N=7 & N=10  \\
\hline
$1/r_{12}$   &  $ 1.58387059E-04 $ & $ 8.59742086E-06 $ & $ 1.70226807E-04 $ & $ 9.32504058E-06 $  \\
$r_{12}$             &  $ 2.31911797E-05 $ & $ 1.04832261E-06 $ & $ - $ & $ -  $  \\
\hline
Series count   & N=20 & N=30 & N=20 & N=30  \\
\hline
$1/r_{12}$   &  $ 7.10235347E-10 $ & $ 7.17956100E-14 $ & $ 7.82848163E-10 $ & $ 7.97365766E-14 $  \\
$r_{12}$             &  $ 6.63828690E-11 $ & $ 6.04972850E-15 $ & $ - $ & $ -  $  \\
\hline
\end{tabular}
\end{table*}

% Tab.2 %
\begin{table*}
\centering
\caption{ AAD and NAAD near vicinity of singularities and cusp region\\
 AAD and NAAD were averaged through 31 angles $\theta_1 \in \langle 0,\pi \rangle$} 

\begin{tabular}{|l|c|c|}
\hline
 TAB2          &   AAD  & NAAD  \\
\hline
Series count   &  N=30  &  N=30  \\
\hline
$1/r_{12}$   &  $ 6.20562199E-05 $ & $ 2.63539895E-13 $   \\
$r_{12}$             &  $ 1.02031516E-16 $ & $ -  $   \\
\hline
\end{tabular}
\end{table*}

\section{Discussion}
The advantage of the new expansion of the Coulomb potential and the linear $r_{12}$ function is their accuracy, results are close 
to the exact potential also near singularities and close to the exact values of the linear term near cusp points. Its disadvantage is the complexity
of the method. The author found also other methods of expansion of the potential, but the method shown in this article seems to us to be the most promising. 
Future work will be focused on a creation of the integral package for calculations of the integrals, which are necessary 
for quantum problems in the case, where our method for expansion of the Coulomb potential will be used.

\section{Acknowledgement}
The author is deeply indebted to  Ji\v{r}\'{i} Pittner for help with preparing the manuscript and for important valuable discussions
of mathematical aspect of the problem. The autor is grateful for consultation the mathematical problems also to Ilja Marti\v{s}ovit\v{s} and
Roman \v{C}ur\'{i}k. Special thanks are due to \v{S}tefan Varga for his valuable comments on the article, previous version of the method and
for its numerical testing. Author is very grateful \\
to \v{S}tefan Dobi\v{s} for technical support and to Michal Kop\v{c}ok for preparing a figures for the article. \\
I would like to express many thanks to my doctor Lucia Mikul\'{a}\v{s}ov\'{a},
without her care our work would not be finished.
This work is dedicated to the memory of my father J\'{a}n Habrovsk\'{y}.\\
Our research was supported by the grant  GACR 19-01897S.

\end{document}